\documentclass[english,preprint,showpacs,aps]{revtex4}
\usepackage[T1]{fontenc}
\usepackage[latin1]{inputenc}
\usepackage{amsmath}
\usepackage{graphicx}
\usepackage{amssymb}
\makeatletter
\usepackage{graphicx}
\usepackage{graphicx,amsfonts}% Include figure files
\usepackage{dcolumn}% Align table columns on decimal point
\usepackage{bm}% bold math
\usepackage{babel}
\makeatother
\begin{document}

\title{The ($2+1$)D Noncommutative CP$^{N-1}$ Model}

\author{E. A. Asano, M. Gomes, A. G. Rodrigues  and A. J. da Silva}

\affiliation{Instituto de F\'{\i}sica, Universidade de S\~{a}o Paulo\\
 Caixa Postal 66318, 05315-970, S\~{a}o Paulo - SP, Brazil}

\email{asano,mgomes,alexgr,ajsilva@fma.if.usp.br}
\date{\today}

\begin{abstract}
We investigate possible extensions of the (2+1) dimensional $CP^{N-1}$
model to the noncommutative space. Up to the leading nontrivial order
of $1/N$, we prove that the model restricted to the left fundamental
representation of the gauge group is renormalizable and does not have
dangerous infrared divergences. In contrast, if the basic field
$\varphi$ transforms in accord with the adjoint representation,
infrared singularities are present in the two point function of the
auxiliary gauge field and also in the leading correction to the
self-energy of the $\varphi$ field. These infrared divergences
may produce nonintegrable singularities leading at higher orders to a
breakdown of the $1/N$ expansion.  Gauge invariance of the
renormalization procedure is also discussed.
\end{abstract}

\maketitle

\section{INTRODUCTION}

One of the main characteristics of field theories defined in noncommutative
space is the infrared/ultraviolet (IR/UV) mixing, which, even in models
without massless particles, leads to the appearance of infrared divergences
and, as a consequence, to the breakdown of the perturbative scheme
in many renormalizable models (see \cite{Douglas} for recent reviews).

The presence of infrared divergences in ordinary field theory signals
that one may be expanding around a point of nonanaliticity of the
exact solution. It may indicate the existence of nonperturbative
effects that can not be reached by a power series expansion on the
perturbative coupling. In such case, two possible approaches are
envisaged. One may try resummations to rearrange the perturbative
series to get a better behaved expansion. A difficulty in this method
is the identification of a parameter to control different orders of
the new series. Another possible procedure is to enlarge the theory
with new interactions, which, hopefully, will cancel the IR
divergences leading to a new expansion without the mentioned
singularities. For noncommutative theories both methods have been considered in the literature
\cite{Minwalla,Griguolo,Bichl,Gomes,Girotti,susskind,Zanon}.  In fact, it has
been argued that the resummation may be efficiently controlled by the
Wilsonian renormalization group, a la Polchinski \cite{Polchinski}. On
the other side, it has been shown that there exists a special class of
theories, namely supersymmetric models, which are natural candidates
to be consistent on noncommutative space, at least as far
renormalization is concerned. This has been proved to be correct for
the noncommutative versions of the four dimensional Wess-Zumino model
\cite{Gomes,Bichl} and the three dimensional nonlinear sigma model
\cite{Girotti} to all orders and also, at least up to one-loop order,
for some supersymmetric gauge models \cite{susskind,Zanon}.  However,
noncommutative theories are so subtle and unusual that detailed
investigations even in nonsupersymmetric theories are still in order.

Proceeding with the aforementioned investigations here we will study
the noncommutative
$CP^{N-1}$ model. 
In this model
local gauge invariance is attained through a composite field that,
at least classically, is not dynamical. This simplifying aspect makes
the model a good laboratory for the investigation of general properties
of gauge fields in noncommutative space. The introduction of a gauge
symmetry in noncommutative space produces a very rich structure in
the sense that, even for the $U(1)$ gauge group, there are three
alternative ways in which the basic matter field could transform.

We begin our study by considering here the pure $CP^{N-1}$ model,
i. e., without  fermionic matter fields. As it happens with its
real $O(N)$ symmetric counterpart, the nonlinear sigma model, only in
two dimensions the commutative version of the model is perturbatively
renormalizable.  However, both in two and three space-time dimensions
it is $1/N$ expandable \cite{Adda,Arefeva}. Dynamical generation of gauge degrees of
freedom and confinement are interesting aspects of the $1/N$ expansion
of the two dimensional model \cite{Adda}. When coupled to fermions
either minimally or in a supersymmetric fashion the quanta of the
basic field $\varphi $ are liberated and exact $S$ matrices are found
\cite{kurak}.

The three dimensional model also possesses some interesting properties.
Its $1/N$ expansion
presents phases in which the basic fields are either massive or
massless \cite{Arefeva}. In particular, if a Chern-Simons term is
added \cite{park,hong,ferretti} one finds radiative corrections to
the topological mass at the next to leading order of $1/N$ \cite{park}.
In this { study} we will work in the unbroken phase (massive $\varphi$)
of the  2+1 dimensional model.

In the noncommutative $CP^{N-1}$ model, because of the underlying
noncommutativity, we may consider the basic field as belonging
alternatively to a fundamental (left or right) or to the adjoint
representation of the gauge group.  We present a detailed discussion
of the renormalization of the model in the fundamental representation
up to the next to leading order of $1/N$. The model turns out to be renormalizable but the
existence of planar and nonplanar graphs with distinct UV behaviors
unveils some interesting features. In particular, some graphs in the
commutative case, as a consequence of charge conjugation do not
contribute. However, these graphs in the noncommutative setting, where charge
conjugation no longer holds \cite{Sheikh}, produces non null
results. In spite of that,  at least up to the
leading nontrivial order of $1/N$, the model turns out to be
renormalizable and free of dangerous infrared divergences \cite{footnote1}.

In contradistinction to the left fundamental representation, the
adjoint representation, already at leading order, presents infrared
singularities. The implications of   these singularities are twofold.
On one hand, those divergences that occur in the gauge sector suggest
the existence of strong long range forces. 
Besides that, in the $\varphi$ field self-energy
corrections there are also quadratic infrared divergences which at
higher order will destroy the $1/N$ expansion. It could be argued
that, similarly to noncommutative $QED_4$ \cite{Hayakawa}, this
behavior may be ameliorated by the inclusion of fermionic fields. This
will be the subject of a subsequent paper where we will discuss the
dynamical generation of a Chern-Simons term. The  elimination
of the (dangerous) IR/UV mixing in a 
 supersymmetric extension of the model  will be also 
investigated.

This work is organized as follows. In Sec. II the possible
representations for the noncommutative $CP^{N-1}$ model are presented.
In Sec. III we investigate the leading contributions to the case in
which the basic field belongs to the left fundamental representation
and proceed a detailed examination of both the UV and IR divergences
up to the next to  leading order of $1/N$. Dimensional
regularization is used and we prove that the model is free from
dangerous divergences (i. e.,  nonrenormalizable or nonintegrable IR
divergences are absent).  In Sec. IV we analyze the behavior of the
Green functions when the basic fields belong to the adjoint representation. In
this situation we explicitly verify the presence of IR/UV mixing which
jeopardizes the consistency of the model.  In Sec.  V we present some
concluding remarks. In the Appendix we discuss some additional
properties of the model.

\section{The Noncommutative CP$^{N-1}$ model}

The commutative CP$^{N-1}$ model is specified by the Lagrangian density

\begin{equation}
{\mathcal{L}}=(D_{\mu }\varphi )^{\dag }D^{\mu }\varphi -m^{2}\varphi ^{\dag }\varphi +\lambda \left (\varphi ^{\dag }\varphi -\frac{N}{g}\right ),\label{1}
\end{equation}

\noindent 
where $\varphi _{i},\quad i=1,\ldots ,N$, are complex scalar fields,
$D_{\mu }\varphi \equiv(\partial _{\mu }+iA_{\mu })\varphi $ is the
covariant derivative of $\varphi $ and  $A_{\mu }$ an auxiliary gauge
field (classically it is just a convenient notation for the composite
field $\frac{g}N(\varphi ^{\dag }\stackrel{\leftrightarrow } {\partial _{\mu
}}\varphi )$); $\lambda $ is the Lagrange multiplier field enforcing
the constraint $\varphi ^{\dag }\varphi =N/g$. Because of this
constraint, the presence of the mass term is classically not
relevant. At the quantum level, $m$ will be identified with the
physical mass for the quanta of the $\varphi $ field insofar one
enforces zero vacuum expectation value for the $\lambda $ field. The
discussion of this fact is entirely analogous to the one in the $O(N)$
nonlinear sigma model and will not be pursued here \cite{Girotti}. The
noncommutative versions of the model are obtained by replacing the
ordinary pointwise product by the  Moyal product \cite{Weyl,Moyal} which is associative
and satisfies \cite{Filk}

\begin{equation}
\phi _{1}(x)\ast \phi _{2}(x)=\lim _{y\rightarrow x}{\textrm{e}}^{\frac{i}{2}\Theta ^{\mu \nu }\frac{\partial \phantom a}{\partial y^{\mu }}\frac{\partial \phantom b}{\partial x^{\nu }}}\phi _{1}(y)\phi _{2}(x),
\label{2}\end{equation}

\noindent
where the constant and antisymmetric matrix $\Theta_{\mu\nu}$ gives a measure
of  the
noncommutativity strength. To evade possible unitarity and/or causality
problems \cite{Gomis1} we will keep $\Theta_{0i}=0$ (see also Ref. \cite{Gomis2}).

As the Moyal ordered product is noncommutative, we shall investigate
three possible representations for the matter field:

1. Left representation:

\begin{eqnarray}
\varphi  & \rightarrow  & ({\textrm{e}}^{i\Lambda })_{*}*\varphi, \\
\varphi ^{\dag } & \rightarrow  & \varphi ^{\dag }*({\mathrm{e}}^{-i\Lambda })_{*},\label{n1}
\end{eqnarray}

where $\Lambda $ is the gauge transformation function and

\begin{equation}
({\textrm{e}}^{i\Lambda })_{\ast }\equiv 1+i\Lambda +\frac{i^{2}}{2}\Lambda \ast 
\Lambda +\ldots \label{3}
\end{equation}

2. Right representation:

\begin{eqnarray}
\varphi  & \rightarrow  & \varphi *({\textrm{e}}^{-i\Lambda })_{\ast },\\
\varphi ^{\dag } & \rightarrow  & ({\textrm{e}}^{i\Lambda })_{*}*
\varphi ^{\dag } .\label{n2}
\end{eqnarray}

3. Adjoint representation:

\begin{eqnarray}
\varphi  & \rightarrow  & ({\textrm{e}}^{i\Lambda })_{\ast }\ast \varphi \ast ({\textrm{e}}^{-i\Lambda })_{\ast },\\
\varphi ^{\dag } & \rightarrow  & ({\textrm{e}}^{i\Lambda })_{\ast }\ast \varphi ^{\dag }\ast ({\textrm{e}}^{-i\Lambda })_{\ast }\label{n3}.
\end{eqnarray}

To keep the action unchanged under these transformations, the usual
derivatives are replaced by covariant derivatives defined as

\begin{eqnarray}
D_{\mu }\varphi =\partial _{\mu }\varphi +iA_{\mu }\ast \varphi \qquad  &  & 
\mbox {left representation,}\label{n4}\\
D_{\mu }\varphi =\partial _{\mu }\varphi -i\varphi \ast A_{\mu }\qquad  &  & 
\mbox {right representation, and}\label{n5}\\
D_{\mu }\varphi =\partial _{\mu }\varphi +iA_{\mu }\ast \varphi -i\varphi 
\ast A_{\mu }\qquad  &  & \mbox {adjoint representation}.\label{n6}
\end{eqnarray}

\noindent 
In all three above representations, the gauge field transforms
as

\begin{equation}
A_{\mu }\rightarrow ({\textrm{e}}^{i\Lambda })_{\ast }\ast A_{\mu }\ast 
({\textrm{e}}^{-i\Lambda })_{\ast }+i[\partial _{\mu }
({\textrm{e}}^{i\Lambda })_{\ast }]\ast ({\textrm{e}}^{-i\Lambda })_{\ast }.
\label{4}
\end{equation}

For sake of simplicity, we shall restrict our analysis to the left
and adjoint representations, as the analysis for right and left representations
are very similar. In the left representation the part of the Lagrangian
containing the auxiliary field $\lambda $ must be written either
as

\begin{equation}
\lambda \ast (\varphi ^{\dag }*\varphi -\frac{N}{g}),\label{5}\end{equation}

\noindent if $\lambda $ does not change or

\begin{equation}
\lambda \ast (\varphi \ast \varphi ^{\dag }-\frac{N}{g}),\label{6}
\end{equation}

\noindent if $\lambda $ changes according to the adjoint representation.

If $\varphi $ belongs to the adjoint representation then $\lambda $
also belongs to this representation and the constraint part of the
Lagrangian should be of the form

\begin{equation}
\lambda *(a\varphi ^{\dag }\ast \varphi +b\varphi \ast \varphi ^{\dag }
-\frac{N}{g}),\label{7}
\end{equation}

\noindent where $a$ and $b$ are free parameters. In what follows,
no matter what representation for the $\varphi $ field is adopted, we always
assume that $\lambda $ belongs to the adjoint representation. As shall
be clear in the next section, a great advantage of this assignment
is the independence of the $\lambda$ and $A_\mu$ fields in the fundamental
representation  (at the leading order 
of $1/N$).      
With this choice, the noncommutative action for the $CP^{N-1}$ model in the
left representation reads

\begin{equation}
{\mathcal{L}}=(D_{\mu }\varphi )^{\dag }*D^{\mu }\varphi -m^{2}
\varphi ^{\dag }*\varphi +\lambda *(\varphi *\varphi ^{\dag }
-\frac{N}{g}).\label{1a}
\end{equation}

As we will do shortly, to  complete this Lagrangian we shall add to it a 
gauge fixing and Faddeev-Popov terms.

\section{The noncommutative $CP^{N-1}$ model in the left representation}

\begin{figure}
\begin{center}\includegraphics{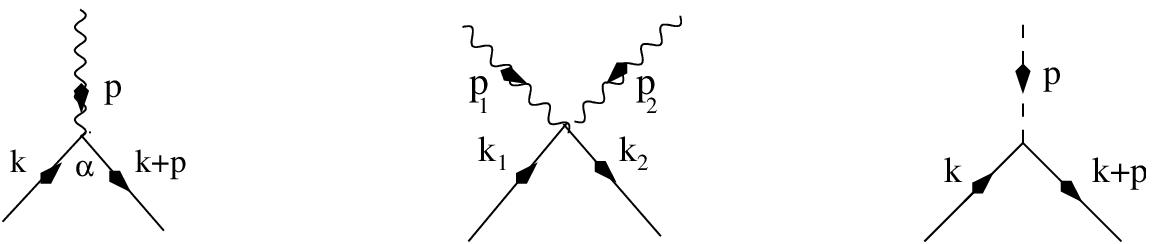}
\end{center}
\caption{Interaction vertices associated to the Lagrangian (\ref{1a}). 
The propagators
for the $A_{\mu }$, $\lambda $ and $\varphi $ fields are represented
by wavy, dashed and continuous lines, respectively. For the complex
field, charge flows in the opposite direction to the indicated.}
\label{Fig1}
\end{figure}

For the left fundamental representation our graphical notation prescribes
the following Feynman rules:

\begin{equation}
\Delta _{\varphi }(p)=\frac{i}{p^{2}-m^{2}+i0},\label{8}\end{equation}

\noindent for the $\varphi $ propagator and 

\begin{eqnarray}
 &  & iA^{\alpha }(\varphi \partial _{\alpha }\varphi ^{\dag }-\partial _{\alpha }\varphi \varphi ^{\dag })\quad \mbox {vertex}\qquad \leftrightarrow \qquad -i(2k+p)_{\alpha }{\textrm{e}}^{-ik\wedge p}\label{n7}\\
 &  & A_\mu A_\nu\varphi \varphi ^{\dag }\quad \mbox {vertex}\qquad \qquad \qquad \qquad \phantom {i}\leftrightarrow \qquad 2ig_{\mu\nu}{\textrm{e}}^{-ik_{1}\wedge k_{2}}
\cos (p_{1}\wedge p_{2})\label{n8}\\
 &  & \lambda \varphi \varphi ^{\dag }\quad \mbox {vertex}\phantom {ab}
\qquad \qquad \qquad \qquad \leftrightarrow \qquad i{\textrm{e}}^{-ik\wedge p}
\label{n9}
\end{eqnarray}

\noindent for the vertices (see Fig. \ref{Fig1}), where $a\wedge
b\equiv \frac{1}{2}a^{\mu }b^{\nu }\Theta _{\mu \nu }$. Except for some
graphs containing the quadrilinear vertex (\ref{n8}), in the
left representation, new features associated with the noncommutativity
are present only for graphs with more than two vertices. This fact depends
crucially on our choice for the $\lambda$ field as belonging to the
adjoint representation, which fixes the sign of the phase in (\ref{n9}).
In particular,
the leading $1/N$ contribution for the mixed propagator $<T\lambda A_\mu>$
is the same as in commutative situation and therefore vanishes,
due to Lorentz covariance.

\begin{figure}
\begin{center}
\includegraphics{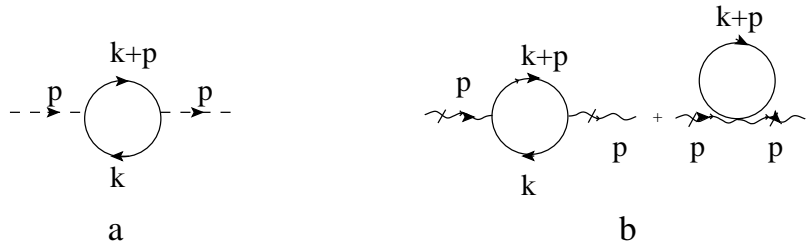}
\end{center}
\caption{Diagrams contributing to the proper functions of the $\lambda$ and
$A_\mu$ fields.}
\label{Fig1.0}
\end{figure}

Contrarily to the $O(N)$ nonlinear sigma model, we will demonstrate
that it is possible to construct a renormalizable model without 
nonintegrable IR/UV mixing.  Actually, we have:

a. $\lambda $ field propagator: $\Delta _{\lambda }(p)=
-1/{F_{\lambda }(p)}$
where (see Fig. \ref{Fig1.0}a)

\begin{equation}
F_{\lambda }(p)=N\int \frac{d^{3}k}{(2\pi )^{3}}\frac{1}{(k+p)^{2}-m^{2}}
\frac{1}{k^{2}-m^{2}}
\approx \frac{iN}{8\sqrt{-p^{2}}}(1-\frac{4m}{\pi}\frac{1}
{\sqrt{-p^ 2}}),\label{9}
\end{equation}

\noindent 
and the expression in the right corresponds to the large spacelike $p$
behavior of $F_{\lambda }(p)$; as shown in the Appendix, for the
analysis of the renormalization of the theory only the leading
${1}/{\sqrt{-p^ 2}}$ is relevant. It should be remarked also that the
above propagator does not have poles and therefore does not have a
particle content.

b. Gauge field two point proper function (Fig. \ref{Fig1.0}b):

\begin{equation}
F_{\mu \nu }(p)  =  N\int \frac{d^{3}k}{(2\pi )^{3}}\left\{ 
\frac{(2k+p)_{\mu }(2k+p)_{\nu }}{[(k+p)^{2}-m^{2}](k^{2}-m^{2})}-
\frac{2g_{\mu \nu }}{k^{2}-m^{2}}\right\} , 
\end{equation}

\noindent 
which turns out to be finite if a gauge invariant regularization
is adopted. Indeed, using dimensional regularization, we obtain

\begin{equation}
F_{\mu\nu}(p)=-\frac{iN}{8\pi }(g_{\mu \nu }-\frac{p_{\mu }p_{\nu }}{p^{2}})p^2
F(p) ,\label{10}
\end{equation}

\noindent where $F(p)=\int
_{0}^{1}dx\frac{(1-2x)^{2}}{[m^{2}-p^{2}x(1-x)]^{1/2}}$. Differently from
 the $\lambda$ field, the gauge field has a
particle interpretation. Indeed, $F(0)= {1}/({3 m})$ so that for
small momenta $A_\mu$ behaves as a Maxwell field of intensity $\frac12
\sqrt{\frac{N}{3\pi m}}$ times the usual one.  For large spacelike momenta

\begin{equation}
F_{\mu\nu}(p)
\approx   \frac{iN}{8\pi}(g_{\mu \nu }-\frac{p_{\mu }p_{\nu }}{p^{2}})
(\frac{\pi}{2}\sqrt{-p^{2}}- 2m).\label{51}
\end{equation}

To get the propagator from (\ref{10}) it is necessary
to fix the gauge. We choose to work in the Landau gauge by adding
to the Lagrangian (\ref{1a}) the term

\begin{equation}
-\frac{N}{2\alpha }(\partial _{\mu }A^{\mu })*(\partial _{\nu }A^{\nu })
+N\partial _{\mu }\overline{C}*[\partial ^{\mu }C+i(C*A^{\mu }
-A^{\mu }*C)]\end{equation}

\noindent and letting $\alpha \rightarrow 0$ after the calculation. Notice the presence of
the Faddeev-Popov ghost term, which due to the non Abelian character
of the Moyal product does not decouple (the ghost fields will not show
up in our leading order calculations but will be relevant in higher
orders). It is now straightforward to verify that the gauge field
propagator is given by

\begin{equation}
\Delta _{\mu \nu }(p)=-\frac{8\pi i}{N}(g_{\mu \nu }-\frac{p_{\mu }p_{\nu }}
{p^{2}})\frac{1}{p^{2}F(p)}
\approx\frac{16 i}{N}(g_{\mu \nu }-
\frac{p_{\mu }p_{\nu }}{p^{2}})(\frac{1}{\sqrt{-p^{2}}}-\frac{4m}{\pi p^2}) .
\end{equation}

As a last remark on the Feynman rules notice that, as in the commutative
theory, any graph containing the diagrams of Fig. \ref{Fig1.0} as subgraphs must
be omitted since those (sub) graphs were already  considered (to
construct the propagators for the $A_\mu$ and $\lambda$ fields).

With these results at hand, we determine the ultraviolet degree of
superficial divergence for a generic graph $\gamma $ as being

\begin{equation}
d(\gamma )=3-N_{A}-2N_{\lambda }-\frac{N_{\varphi }}{2}-\frac{N_C}2.\end{equation}

\noindent 
where $N_{A}$, $N_{\varphi }, N_{\lambda }$ and $N_C$ are
the number of the external lines associated to the gauge, $\varphi $,
 $\lambda $  and ghost fields, respectively.  Renormalization parts are those
graphs having $d(\gamma)\geq 0$; in a noncommutative theory they occur only
for planar (sub) graphs. 
Some of the ultraviolet divergences, associated with the planar graphs,
may be absorbed by reparametrizations.  
As usual, we define the renormalized quantities by the replacements

\begin{eqnarray}
A_\mu &\rightarrow&  Z^{1/2}_{A}A_\mu=(1+a) A_\mu,\\
\varphi &\rightarrow & Z^{1/2}_{\varphi}\varphi= (1+b)^{1/2} \varphi,\\
\lambda &\rightarrow & Z^{1/2}_{\lambda}\lambda= (1+c)\lambda,\\
1/g &\rightarrow & Z_g /g= (1+d)/g,
\end{eqnarray}

\noindent
so that the Lagrangian (\ref{1a}) written in terms of the new fields
changes as ${\mathcal L}\rightarrow {\mathcal L} +{\mathcal L}_{ct}$, where
the counterterm Lagrangian is given by

\begin{eqnarray}
{\mathcal{L}}_{ct}&=&b\,\partial_{\mu }\varphi^{\dag }*\partial^{\mu }\varphi
-m^2 b\,\varphi ^{\dag }\varphi +iB(\partial_\mu \varphi^{\dag}*A^\mu *\varphi-
\varphi^{\dag}*A^\mu *\partial_\mu \varphi)\nonumber\\
&\phantom a &+C \varphi^{\dag} *A_\mu *A^\mu * \varphi+
D\lambda *\varphi *\varphi ^{\dag }
-F N\frac{\lambda}g ,\label{count}
\end{eqnarray}

\noindent 
where we introduced
\begin{eqnarray}
B&=&(1+a)(1+b) -1,\\
C&=&(1+a)^2 (1+b) -1,\\
D&=&(1+c)(1+b) -1,\\
F&=& (1+c)(1+d) -1. 
\end{eqnarray}

These counterterms may be used to 
enforce $m$ as the physical mass of the $\varphi$ field, to ensure
the elimination of the remaining divergences of the two point function
of the $\varphi$ field and of the three point function $<TA_{\mu}\varphi^\dag
\varphi>$.

The analysis of the UV divergences is much facilitated by the help of
the graphical identities \cite{Arefeva} depicted in
Fig. \ref{Fig1a}. Due to the independence of the auxiliary field
propagators on the noncommutative parameter, these identities are
valid also in the present situation.  It should be observed that, as
the $\lambda$ field has no particle content, the identities may be
used even if we restrict ourselves to the one particle irreducible
graphs, i. e., to graphs which can not be separated into disjoint
pieces by cutting just one line, wavy or continuous.  Before going any
further we would like to stress some important consequences of these
identities.  As in the commutative case, the identity of Fig. \ref
{Fig1a}a implies that the $\varphi$ mass counterterm is innocuous
since it cancels in all contributions to the Green functions. This
will be explicitly verified in our discussion of the renormalization
of the two point function of the $\varphi $ field. Another implication
of the graphical identity is that the $D\lambda \varphi^\dag \varphi$
counterterm is also innocuous if we consider Green functions of the
$\varphi$ and $A_\mu$ fields only (no external $\lambda$ lines); in
that case $D$ may be chosen at will and the wave function
renormalization for the $\lambda$ field is therefore irrelevant.  In
our approach  the $\lambda$ field tadpole contributions will not be
considered separately but just in connection with the computation of
the two point Green function of the $\varphi$ field. 

An important implication of the identity of Fig. \ref{Fig1a}b  is
that, except for the second diagram of
Fig. \ref{Fig1.0}b, all contributions containing the quadrilinear
vertex will cancel pairwise; they need not be considered anymore.

We also need to consider those divergences which do not have a
corresponding counterterm. They may have $N_{\lambda }$ equal to
either $0$ or $1$ . For $N_{\lambda }=1$, the dangerous divergences
are associated with graphs with $N_{\varphi }=0$ and $N_{A}=1$. As
mentioned earlier, this last possibility does not happen if a Lorentz
covariant regularization is employed.

For $N_{\lambda }=0$ there are more possibilities:

1. Graphs with $N_{A}=0$ and $N_{\varphi }$ equal to either $4$
or $6$,

2. Graphs with $N_{A}=1$ and $N_{\varphi }=4$,

3. Graphs with $N_{A}=2$ and $N_{\varphi }=0$,

4. Graphs with $N_{A}=3$ and $N_{\varphi }=0$.

Besides the UV behavior, in all cases we need to investigate the
possible presence of infrared divergences (UV/IR mixing).

We focus first on the processes whose corresponding
counterterms are correlated by gauge invariance, namely, corrections
for the $\varphi$ propagator, the three point $<TA_\mu\varphi
\varphi^{\dagger}>$ and the four point $<TA_\mu A^\mu
\varphi^{\dagger} \varphi>$ functions. We have:

1. The subleading contributions to the self-energy of the $\varphi $
field, $\Gamma(p)$, are shown in Figs. \ref{Fig2} and
\ref{Fig2a}. They are purely planar and their (ultraviolet)
divergences should be absorbed into a mass and wave function
counterterms for the $\varphi $ field. The mass counterterm is
associated to the highest (quadratic) divergence gotten by setting
zero the external momentum of the contributing graphs.  As can be
easily checked, these divergences cancel between Figs. \ref{Fig2}a and
\ref{Fig2}b, due to the graphical identity of Fig.~\ref{Fig1a}a.

The
contributions for the wave function renormalization of the $\varphi$
field come from Figs. \ref{Fig2a}a and 
\ref{Fig2a}b. Using dimensional regularization, a straightforward 
calculation  furnishes the following results

\begin{eqnarray} 
\Sigma_{\varphi}^{(a)}(p)&=&-i\int\frac{d^Dk}{(2\pi)^D}\frac{(k+2p)_{\mu}
(k+2p)_{\nu}}{(k+p)^2-m^2}\Delta^{\mu\nu}(k)=-i\frac{1}{N}\frac{64 p^2}{3\pi^2
\epsilon}
+\mbox{ finite terms}\\
\Sigma_{\varphi}^{(b)}(p)&=&-i\int\frac{d^D k}{(2\pi)^D}\frac{1}{(k+p)^2-m^2}
\Delta_{\lambda}(k)=C_b+i\frac{1}{N}\frac{4p^2}{3\pi^2\epsilon}+
\mbox{ finite terms},
\end{eqnarray}

\noindent
where $\epsilon=D-3$ and $C_b$ is a quadratically divergent
constant that would contribute to the mass renormalization of the
$\varphi$ field; as mentioned the mass renormalization terms cancel.
Diagram \ref{Fig2a}c, on the other hand, cancels between
Figs. \ref{Fig2}a and \ref{Fig2}b due to Fig. \ref{Fig1a}b. The
divergent parts are therefore eliminated by the counterterm
$\frac{1}{N}\frac{20}{\pi^2\epsilon}\partial_\mu\varphi^{\dagger}
\partial^\mu\varphi$ which fixes the divergent part of $b$ as being
$b_{div}=\frac{1}{N}\frac{20}{\pi^2\epsilon}$. The overall divergences
associated with the tadpole in Fig. \ref{Fig2}b are absorbed in the
counterterm in Fig.~\ref{Fig2}d.  We also assume that $F$
possesses a finite part which enforces $m$ as the physical mass,
i. e., by adjusting $F$ and $b$ we impose the following normalization
conditions

\begin{eqnarray}
\Gamma(p)&=& 0\qquad \mbox{for $p^2 =m^2$ and}\\
\frac{\partial \Gamma}{\partial p^2}&=& 0\qquad \mbox{for $p^2 =m^2$}.
\end{eqnarray}

2. Three point function of the $A_{\mu }$ and $\varphi $ fields,
i. e., $<TA_{\mu }\varphi ^{\dag }\varphi >$.  Because of our previous
remark on the cancellation of diagrams containing the quadrilinear
vertex $A_{\mu }A^{\mu }\varphi ^{\dag }\varphi $, we have to analyze
only those diagrams without this vertex, i. e., those which are
depicted in Figs.  \ref{Fig3a} and \ref{Fig3}. In Fig. \ref{Fig3}
there are two one-loop diagrams and eight two-loop diagrams.  Notice
that the last four two-loop diagrams differ from the first four
two-loop ones just by the orientation of the charge flow in the upper
bosonic loop.  In the commutative situation, graphs which differ just
by the orientation of the charge flow are related by charge
conjugation and Furry's theorem states that they either give equal
contributions or cancel between themselves. Here however charge
conjugation is lost and Furry's theorem is no longer valid so that the
contributions should be individually analyzed. In the construction of
the diagrams implicit in Fig. \ref{Fig3a}b it is important to notice
that any planar contribution is automatically overall ultraviolet 
finite. Indeed, these planar diagrams have zero degree of superficial
divergence but, because of Lorentz covariance they are proportional to
$p_\mu$ what lowers the efective degree of divergence by one unit. On
the other hand, as we will show, the nonplanar contributions in
Fig. \ref{Fig3a}b are used to cancel infrared divergences in the
nonplanar diagrams of Fig. \ref{Fig3}.

The first two diagrams shown in Fig. \ref{Fig3} are purely
nonplanar and therefore are
ultraviolet finite but could originate nonintegrable (linear) IR
divergences. In fact, because of the transversality of the $\Delta
_{\rho \sigma }$ propagator, the graph \ref{Fig3}a is finite being
given by

\begin{eqnarray}
\int \frac{d^{3}k}{(2\pi )^{3}}\, {\textrm{e}}^{-i(2k\wedge p-p\wedge p_{1})}
\frac{[2(k+p_{1})+p]_{\mu }\, [2(p_{1}+p)]_{\rho }\, 2p_{1\, \sigma }}
{[(k+p_{1})^{2}-m^{2}]\, [(k+p_{1}+p)^{2}-m^{2}]}\Delta ^{\rho \sigma }(k). 
&  & 
\end{eqnarray}

\noindent
Due to the asymptotic behavior of $\Delta _{\rho \sigma }(k)$, this
integral is finite even when the phase factor is absent so that the
result is free from IR singularities.  Graph \ref{Fig3}b, on the
other hand, is linearly divergent at $p=0$. To see how this divergence
is canceled we write its amplitude as

\begin{eqnarray}
\int \frac{d^{3}k}{(2\pi )^{3}}\, {\textrm{e}}^{-i(2k\wedge p
-p\wedge p_{1})}I_{\mu }(k,p,p_{1}) & = & 
\int \frac{d^{3}k}{(2\pi )^{3}}\, {\textrm{e}}^{-i(2k\wedge p
-p\wedge p_{1})}[I_{\mu }(k,0,0)+R_{\mu }(k,p,p_{1})], \label{n12}
\end{eqnarray}

\noindent where

\begin{equation}
I_{\mu }(k,p,p_{1})=\frac{[2(k+p_{1})+p]_{\mu }}{[(k+p_{1})^{2}-m^{2}]\, 
[(k+p_{1}+p)^{2}-m^{2}]}\Delta _{\lambda }(k)
\end{equation}

\noindent and $R_{\mu }(k,p,p_{1})$ presents at most logarithmic
divergences. Explicitly,

\begin{equation}
\int \frac{d^{3}k}{(2\pi )^{3}}\, {\textrm{e}}^{-i(2k\wedge p
-p\wedge p_{1})}\, I_{\mu }(k,0,0)=\mbox {Const.}\, \frac{\tilde{p}_{\mu }}
{\tilde{p}^{2}}{\textrm{e}}^{ip\wedge p_{1}},
\end{equation}

\noindent 
where we introduced a simplified notation $\tilde{p}^{\mu }=\Theta
^{\mu \nu }p_{\nu }$.  Now, among the diagrams implicit in
Fig. \ref{Fig3a}b we consider the diagram of Fig.~\ref{Fig3b}a which
may be obtained from the graph \ref{Fig3}b by joining its $\varphi $
external lines at a new $\lambda \varphi\varphi^\dag$ vertex and
attaching the external lines to a second $\lambda \varphi \varphi^\dag$
vertex linked to the first one by the $\lambda$ propagator. The
amplitude for this graph reads

\begin{equation}
J_\mu(p,p_1)=\Delta _{\lambda }(p)\int \frac{d^{3}k}{(2\pi )^{3}}\frac{d^{3}q}
{(2\pi )^{3}}
{\textrm{e}}^{-i(2k\wedge p-p\wedge p_{1})}I_{\mu }(k,p,q)\frac{i^{2}}{(q^{2}
-m^{2})\, [(q+p)^{2}-m^{2}]}.
\end{equation}

\noindent 
Expanding $I_{\mu }(k,p,q)$ around $p=q=0$ as before and using (\ref{9})
we get

\begin{eqnarray}
J_\mu(p,p_1)&=&-\mbox {Const.}\, \frac{\tilde{p}_{\mu }}{\tilde{p}^{2}}
{\textrm{e}}^{ip\wedge p_{1}}\nonumber\\
&&+\int \frac{d^{3}k}{(2\pi )^{3}}
\frac{d^{3}q}{(2\pi )^{3}}{\textrm{e}}^{-i(2k\wedge p-p\wedge p_{1})}
R_{\mu }(k,p,q)\frac{i^{2}}{(q^{2}-m^{2})\, [(q+p)^{2}-m^{2}]}
\Delta _{\lambda }(p),\label{const}
\end{eqnarray}

\noindent 
where the second term has at most logarithmic IR divergences.
Thus, adding the two contributions, no dangerous IR divergence
survives.  This cancellation is just a manifestation of the identity
expressed in Fig. \ref{Fig1a}a.  Being nonplanar diagrams
\ref{Fig3}a and \ref{Fig3}b do not present ultraviolet divergences
either.  This is an interesting point since in the renormalization of
commutative $QED$ model the contribution of the diagram \ref{Fig3}a
is  important to secure the gauge invariance of the perturbative
method. The above procedure can be generalized for any linearly IR
divergent graph. From any nonplanar graph $\gamma$ we may construct a
new diagram $\bar \gamma$ by joining two external $\varphi$ lines of
$\gamma$ in a new trilinear vertex $\lambda \varphi
\varphi^\dag$. This new diagram contains $\gamma$ as a subgraph so
that it presents the same IR divergence as $\gamma$. The divergence in
$\bar \gamma$ may be extracted by a simple Taylor expansion as we did
in the above calculation.  Summing the analytical expressions for
$\gamma$ and $\bar\gamma$ it remains only a mild logarithmic
IR divergence.

The next set of  graphs shown in Fig. \ref{Fig3} consist of four planar
diagrams, Figs. \ref{Fig3}(c-f). As those
graphs have two loops they may have one-loop divergent subgraphs.
Here, however, we are concerned only with the overall divergence
postponing the analysis of the divergences of the subgraphs to a later
discussion (see Appendix).  Graph \ref{Fig3}c is actually finite since 
due to the
transversality of the gauge field propagator the external vertices in
the lower line can not depend on the loop momentum containing the two
wavy lines. In the sequel we list the divergent contributions arising
from the diagrams \ref{Fig3}(d-f):

a. Graph \ref{Fig3}d 
 
\begin{eqnarray}
\Gamma^{(1,2)}_{\mu(d)}&=&-i4Ne^{-ip_1\wedge p}\int
 \frac{d^Dk}{(2\pi)^D}\frac{d^Dq}{(2\pi)^D}\frac{(2q+p)_\mu q_\beta
 p_{1\alpha}\Delta^{\alpha\beta}(k)\Delta_\lambda(k-p)}{[(k+p_1)^2-m^2][(k+q)^2
-m^2](q^2-m^2)[(q+p)^2-m^2]}\nonumber\\
&= &i(p_1)_\mu e^{-ip_1\wedge p}\frac{16}{3N\pi^2}\frac{1}{\epsilon}+ 
\mbox{finite terms}.
\end{eqnarray}

b. Graph \ref{Fig3}e

\begin{eqnarray}
\Gamma^{(1,2)}_{\mu(e)}&=&-iNe^{-ip_1\wedge
  p}\int\frac{d^Dk}{(2\pi)^D}\frac{d^Dq}{(2\pi)^D}\frac{(2q+p)_\mu
(k+2q+2p)_\beta(k+2p+2p_1)_\alpha}{[(k+p+p_1)^2-m^2][(k+q+p)^2-m^2]}
\nonumber\\
& &\times
  \frac{\Delta^{\alpha\beta}(k)\Delta_\lambda(k+p)}{(q^2-m^2)[(q+p)^2-m^2]}
 \nonumber\\
&=& i(p_1+p)_\mu e^{-ip_1\wedge p}\frac{16}{3N\pi^2}\frac{1}{\epsilon}+ 
\mbox{finite terms}.
\end{eqnarray}

c. Graph \ref{Fig3}f

\begin{eqnarray}
\Gamma^{(1,2)}_{\mu(f)}&=&-iNe^{-ip_1\wedge
  p}\int\frac{d^Dk}{(2\pi)^D}\frac{d^Dq}{(2\pi)^D}\frac{(2q+p)_\mu
\Delta_\lambda(k)\Delta_\lambda(k-p)}{[(k+p_1)^2-m^2][(k+q)^2-m^2]}\nonumber\\
& & \times \frac{1}{(q^2-m^2)[(q+p)^2-m^2]} \nonumber\\
&=&-i(2p_1+p)_\mu e^{-ip_1\wedge p}\frac{2}{3N\pi^2}\frac{1}{\epsilon}+ 
\mbox{finite terms}.
\end{eqnarray}

So altogether we have 

\begin{equation}
\Gamma^{(1,2)}_{\mu(d)}+\Gamma^{(1,2)}_{\mu(e)}+\Gamma^{(1,2)}_{\mu(f)}=
i(2p_1+p)_\mu e^{-ip_1\wedge p}\frac{14}{3N\pi^2}\frac{1}{\epsilon}+ 
\mbox{finite terms}.
\end{equation}

The
above divergence can be eliminated by the trilinear counterterm
$\frac{14i}{3N\pi^2\epsilon} A^\mu (\varphi\partial_\mu
\varphi^\dagger-\partial_\mu \varphi \varphi^\dagger)$, so that
$B=\frac{14}{3N\pi^2\epsilon}$. 

Let us now consider the four nonplanar diagrams shown in
\ref{Fig3}(g-j).  As we have outlined in our discussion subsequent to
Eq. (\ref{const}), any infrared linear divergence can be eliminated by
adequately combining the graphs.  Nevertheless, in specific situations
there are further additional simplifications. Indeed, we have:

a. Graph \ref{Fig3}g: Because of our gauge choice, there
is no IR divergence associated to that diagram.

b. Graphs \ref{Fig3}(h-i): Again, due to our gauge choice these
diagrams may present only a mild logarithmic IR divergence. This
divergence is canceled by the corresponding diagrams implicit in Fig.
\ref{Fig3a}b.

c. Graph \ref{Fig3}j: The amplitude associated with this
diagram is linearly divergent at zero external momentum.  Here we
apply the aforementioned  construction which produces the graph shown
in Fig. \ref{Fig3b}b (this is another graph implicit in Fig. \ref{Fig3a}b).
 The leading IR divergences of these two diagrams cancels as we
proved earlier.

Our results can be used now to fix the value of $C$ as defined in the
counterterm Lagrangian (\ref{count}). In fact, as $B=
\frac{14}{3N\pi^2\epsilon}$ then
$a=B-b_{div}=-\frac{46}{3N\pi^2\epsilon}$ so that $C=b_{div}+ 2a=-\frac{32}{3N\pi^2 \epsilon}$.
We remark that the only possible contributions of  $C$
would be for the next to leading corrections to the $A_\mu$
propagator; due to the graphical identity in Fig. \ref{Fig1a}b a
nonvanishing $C$ does have none effect up to the order we have been
considering.

3. Four point function of the $A^\mu$ and $\varphi$ fields, $<TA_\mu
A_\nu \varphi^\dag \varphi>$. There are not ultraviolet divergences
because a given graph is either nonplanar or one may find a
``partner'' graph to which the graphical identity in Fig. \ref{Fig1a}a
may be applied. In the last case, the divergence in the original graph
and its partner cancel pairwise. This is consistent with the fact that
for this four point function no counterterm is effective; in fact, the
absence of these divergences may be considered as a test for the
consistency of the calculation.

Although the
renormalized Lagrangian turned out to be gauge invariant both in
commutative and in the noncommutative cases, the mechanism by which
this gauge invariance is achieved is entirely different in the two
situations. Thus, diagrams \ref{Fig3}a and \ref{Fig3}b which are nonplanar are
ultraviolet finite, contrarily to the
commutative case. On the other hand, in the noncommutative setting
Furry's theorem is not valid and so many graph cancellations that hold
in the commutative case are now absent and new contributions arise.

4. Five point function, $<T A_\mu \varphi^{\dag} \varphi\varphi^{\dag} 
\varphi>$. The contributing diagrams are  at most logarithmically divergent.
 In the planar part this divergence can be get by calculating
the regularized amplitude at zero external momenta (after
extracting the phase factors which in this case do not depend
on the internal momentum). Because of Lorentz covariance, it is
clear that the result of this computation vanishes so that no counterterm
is needed.  The possible IR divergence contained in the nonplanar diagrams of this
type can be canceled using a construction similar to the one described
after Eq. (\ref{const}).

5. Three point vertex function of the $A_\mu$ field, $<TA_\mu A_\nu
A_\rho>$. There are just two one-loop graphs which differ only by the
orientation of the charge flow in the loop (each loop consists of three
bosonic lines). These diagrams are both
planar and adding them one gets a  factor depending on the sine
  of the wedge product of
the two external momenta times an integral which is finite by symmetric
integration.

6. Three point vertex function of the $\lambda $ and $\varphi $
fields, $<T\lambda\varphi^\dag\varphi>$. To order $1/N$ the
contributing graphs are depicted in Fig.  \ref{Fig31}. The one-loop
graphs \ref{Fig31}a and \ref{Fig31}b are nonplanar and therefore
ultraviolet finite although in the infrared limit may present a mild
logarithmic divergence. The graph \ref{Fig31}c, on the contrary, is
planar and is ultraviolet logarithmically divergent.  It has an
analytic expression given by

\begin{eqnarray}
\int \frac{d^{3}k}{(2\pi )^{3}}\frac{d^{3}q}{(2\pi )^{3}}
\frac{{\textrm{e}}^{-i(p_{1}\wedge p)}}{[(k+p_{1})^{2}-m^{2}][(k+p_{1}+p)^{2}
-m^{2}][(k-q)^{2}-m^{2}]} &  & \nonumber \\
\frac{1}{(q^{2}-m^{2})}\Delta _{\lambda }(q+p_{1})\Delta _{\lambda }
(q+p_{1}+p). &  & \label{n10}
\end{eqnarray}

Besides presenting an overall logarithmic divergence this integral has
a divergent subintegral, namely, the $q$ integration (this divergence
will be examined in the context of the four point function of the
$\varphi $ field, in the next item). The overall divergence can not be
eliminated through the use of the $D \lambda \varphi \varphi^{\dag} $
counterterm since contributions containing such counterterm are
canceled due to the identity of Fig. \ref{Fig1a}a. However, as
exemplified in the Appendix, the mentioned divergence is irrelevant as
far the Green functions with only external $\varphi$ and $A_\mu$ fields are
concerned.

The analytic expression for the graph
\ref{Fig31}d differs from (\ref{n10}) just by an additional factor
${\textrm{e}}^{-2iq\wedge p}$ and therefore is ultraviolet finite and
has a mild logarithmic divergence when $p$ tends to zero. Notice that
in the commutative situation graphs \ref{Fig31}c and \ref{Fig31}d
would give the same contributions, as a consequence of charge
conjugation invariance. There are other graphs, not shown in
Fig. \ref{Fig31}, which differ from diagrams \ref{Fig31}c and
\ref{Fig31}d just by the replacements, one at each time, of the
internal dashed lines by wavy ones; because of the transversality of
the $A_\mu$ propagator these additional graphs are ultraviolet finite
and without infrared singularities.

7. Concerning the contributions to the four point function
$<T\varphi\varphi\varphi^\dag\varphi^\dag>$ let us first examine the
one-loop diagrams. One sees that there are two types to be considered
as they are depicted in Fig. \ref{Fig4}. Whereas the graphs in the
first row of Fig. \ref{Fig4} are ultraviolet linearly divergent, the
graphs in the second row are nonplanar and therefore ultraviolet
finites; they do not need counterterms. No counterterm is also needed
for the four  graphs of the first row because, as a consequence
of the graphical identity of Fig. \ref{Fig1a}a, there are two-loop
graphs which cancel the mentioned divergences.  For example, the
highest (linear) divergence of
the graph \ref{Fig4}a  is canceled by
the one associated with the graph \ref{Fig5}b.  Graph \ref{Fig5}b
has a subgraph with the same divergence as the graph \ref{Fig4}a. If
we contract this subgraph to a point and use the identity of
Fig. \ref{Fig1a}a we obtain the cancellation of these divergences.
By a similar mechanism the logarithimic divergences which are proportional
to the external momenta of the graph are also cancelled.
The complete cancellation of all ultraviolet divergences can become
complicate as it is illustrated in the Appendix.

8. Six point function, $<T\varphi\varphi\varphi\varphi^\dag\varphi^\dag
\varphi^\dag>$. As before the divergences of the planar
diagrams cancel pairwise by the use of the identity in Fig. \ref{Fig1a}a
whereas the nonplanar graphs could at most develop a logarithmic 
infrared singularity.

The above discussion proves that, up to the leading nontrivial order of $1/N$,
the noncommmutive $CP^{N-1}$ model is renormalizable and without dangerous infrared
singularities if the $\varphi$ field transforms in accord with the
left fundamental representation. 

\section{The noncommutative $CP^{N-1}$ model in the adjoint representation}

Let us now consider the leading $1/N$ contributions when the basic
fields transform in accord with the adjoint representation. We will
adopt the same graphical notation as in the previous section. However, we have new rules:

\vspace{0.5cm}

1. Trilinear $A_{\mu }\varphi ^{\dag }\varphi $ vertex $\leftrightarrow $
$-2(2k+p)_{\mu }\, \sin (k\wedge p)$.

\vspace{0.4cm}

2. Quadrilinear $A^{\mu }A^{\nu }\varphi ^{\dag }\varphi $ vertex
$\leftrightarrow $ $-4i g^{\mu\nu}[\sin (k_{1}\wedge p_{1})
\sin (k_{2}\wedge p_2)+p_1\leftrightarrow p_2]$.

\vspace{0.5cm}

\noindent 
Notice that these interactions are absent in the commutative
limit.

Using these rules we fix the  two point function of
the gauge field as being

\begin{eqnarray}
F_{\mu \nu }(p)=4N &  & \left[\int \frac{d^{3}k}{(2\pi )^{3}}
\frac{(2k+p)_{\mu }(2k+p)_{\nu }}{(k^{2}-m^{2})[(k+p)^{2}-m^{2}]}\sin ^{2}
(k\wedge p)\right.\nonumber \\
 &  & \left.-2g_{\mu \nu }\int \frac{d^{3}k}{(2\pi )^{3}}\frac{1}{k^{2}-m^{2}}
\sin ^{2}(k\wedge p)\right].\label{n11}
\end{eqnarray}

As $\sin ^{2}(k\wedge p)=\frac{1}{2}(1-\cos 2(k\wedge p))$ we get
a planar part which is twice that of the gauge field two point function
in the corresponding commutative theory. Concerning the nonplanar
piece, we perform the standard procedures to obtain

\begin{equation}
F^{np}_{\mu \nu }(p)=-2N\!\int _{0}^{1}\!dx\!\int 
\frac{d^{3}k}{(2\pi )^{3}}\frac{e^{ik\tilde{p}}}{[k^{2}-M^{2}]^{2}}\{4k_{\mu }
k_{\nu }+p_{\mu }p_{\nu }(2x-1)^{2}-2g_{\mu \nu }[k^{2}+p^{2}(x-1)^{2}
-m^{2}]\}\label{10a}
\end{equation}

\noindent where $M^{2}=m^{2}-p^{2}x(1-x)$. Now, using \cite{Gelfand},

\begin{equation}
\int \frac{d^{3}k}{(2\pi )^{3}}\frac{e^{ik_{\alpha }\tilde{p}^{\alpha }}}
{[k^{2}-M^{2}]^{2}}=\frac{i}{(2\pi )^{3/2}\sqrt{4M}}\frac{K_{-1/2}[M
\sqrt{-{\tilde{p}}^{2} }]}{(-{\tilde{p}^{2}} )^{-1/4}}=\frac{i}{8 \pi}
\frac{{\rm e}^{-M\sqrt{-\tilde {p}^2}}}{M},
\label{I-basica-pura}
\end{equation}

\noindent where $K_{\nu}$ is the modified Bessel function of order
$\nu$, one obtains the complete two point function of the gauge field
(for  simplicity we are employing the same notation
used for the counterterms in the previous section; no confusion should
arise since  they refer to distinct  situations)

\begin{eqnarray}
F_{\mu \nu }(p)&=&(g_{\mu \nu }p^{2}-p_{\mu }
p_{\nu })A +\tilde{p}_{\mu}\tilde{p}_{\nu}B + p_\mu p_\nu C,   \label{n11a}
\end{eqnarray}

\noindent
where

\begin{equation}
A=-\frac{Ni}{4\pi }\int _{0}^{1}dx  [\frac{1}{M}(1-2x)^{2}(1
-{\textrm{e}}^{-M\sqrt{-\tilde{p}^{2}}})],
\end{equation}

\begin{equation}
B =\frac{Ni}{\pi{\tilde p}^2 } \int_{0}^{1} dx\left(\frac{1}
{\sqrt{-\tilde{p}^{2}}}+M\right) {\textrm{e}}^{-M\sqrt{-\tilde{p}^{2}}},
\end{equation}

\noindent 
and a gauge fixing term was added. Notice that this result
possess an
infrared singularity at $\tilde{p}=0$.

Concerning the part of the  Lagrangian which depends on the auxiliary field
$\lambda$ we should recall that, as pointed out in (\ref{7}), there is a
two-parameter family of possible interaction terms; some
simplifications occur depending on which form the interaction is chosen.  In
particular, notice that:
 
1. If the interaction term containing $\lambda$ is   
taken as in the previous section, 
the computation of the two point function of the $\lambda $ field  
gives the same result as before  but, differently from the left
representation,  the mixed propagator $<T\lambda A_\mu>$
turns out to be nonvanishing. In fact, 
we find that at the leading order of $1/N$ the two
point function of the $\lambda$ and $A_\mu$ fields is given by

\begin{equation}
\Gamma _{A_{\mu }\, \lambda }(p)=N\int \frac{d^{3}k}{(2\pi )^{3}}\, 
\frac{(2k+p)_{\mu }}{(k^{2}-m^{2})[(k+p)^{2}-m^{2}]}{
\textrm{e}}^{-i2k\wedge p}.
\end{equation}
 
\noindent
In contrast with the commutative model, the above expression does not
vanish and yields

\begin{equation}
\Gamma _{A_{\mu }\, \lambda }(p)=-\frac{N\tilde{p_{\mu }}}{4\pi 
\sqrt{-{\tilde{p}}^{2}}}
\int_{0}^{1} dx {\rm e}^{-M\sqrt{-{\tilde p}^2}}\equiv D\tilde p_\mu.\label{49}
\end{equation}

\noindent
where $D$ is a nonvanishing function of ${\tilde p}^2$ and $p^2$.

2. If the  interaction term is chosen to be as

\begin{equation}
 \lambda\ast(\phi \ast \phi^{\dag}-\phi^{\dag}\ast \phi),\label{30}
\end{equation} 

\noindent
 then, at the leading
order of $1/N$, the mixed propagator vanishes but the 
$\lambda$ field propagator will have a nonplanar contribution. In this case 
the two point function of the $\lambda$ field  will be

\begin{equation}
2{F_\lambda(p)}+ F_{np\,\lambda}(p)\equiv E(p),\label{50}
\end{equation}

\noindent
where $F_\lambda$ was given in (\ref{9}) and the nonplanar part 
$F_{np\,\lambda}$ is

\begin{equation}  
 F_{np\,\lambda}(p)=-\frac{i N}{4\pi}\int_{0}^{1}dx\, \frac{{\rm e}^{-M
\sqrt{-{\tilde p}^2}}}{ M}.
\end{equation}

As another consequence of the choice (\ref{30}), the graphical
identity of Fig. \ref{Fig1a}a is no longer valid. There are much more
contributing diagrams than in the left representation.

We are now in a position which allows us to compute the propagators for the
$A_\mu$ and $\lambda$ fields at leading order of $1/N$. To encompass
the two situations listed above, we shall designate
the two point function of the auxiliary $\lambda$ field by $E(p)$ with the 
understanding that
for the case {\underline 1} $E(p)$ and $D(p)$ are given by (\ref{9}) and (\ref{49}), respectively  whereas for the
case {\underline 2} $E(p)$ is given by (\ref{50}) and $D(p)=0$.
The propagators are then fixed by the inverse of
the matrix which appears in the quadratic part of the Lagrangian. A direct calculation then furnishes

 $A_\mu$ propagator:

\begin{equation}
\Delta^{\mu\nu} =  (g^{\mu\nu} p^2 - p^\mu p^\nu)  \frac{-1}{(p^2)^2\;A}+ b\,
\tilde{p}^{\mu}\tilde{p}^{\nu} -  \frac{p^\mu p^\nu}{(p^2)^2C} ,
\end{equation}

\noindent
where

\begin{equation}
b=\frac{-D^2}{[E(Ap^2+B\tilde{p}^2)
-D^2\tilde{p}^2](Ap^2+B\tilde{p}^2)}
+ \frac{Bp^2}{p^2 A(Ap^2+B\tilde{p}^2)}.
\end{equation}

Mixed propagator:

\begin{equation}
\Delta^\nu(p) \equiv <TA^\nu\lambda >= d\tilde{p}^\nu ,
\end{equation}

\noindent
where

\begin{equation}
d= \frac{D}{E(A p^2+ B{\tilde p}^2)- D^2 {\tilde p}^2}.
\end{equation}

$\lambda$ propagator:

\begin{equation}
\Delta_\lambda(p)= - \frac{1}{E}(1+ dD{\tilde {p}^2}).
\end{equation}

At small momenta $b\simeq \frac{1}{p^2 {(-{\tilde p}^2)^{3/2}}}$ and
$A\simeq \sqrt{-\tilde {p}^2}$ in both situations discriminated above
and $d\simeq {1}/{{\sqrt{-{\tilde p}^2}}}$ for the case {\underline
1}.  Thus, the transversal part of the $A_\mu$ field propagator
diverges badly (as $\frac1{p^2(-{{\tilde p}^2})^{1/2}}$) at small
momentum. In a local model such behavior would in a nonrelativistic
limit be associated with a potential which grows linearly with the
distance from a charge probe. Therefore the quanta of the $\varphi$
field would be confined. However, due to the nonlocal character of the
interaction, there are at the vertices  momentum dependent form factors 
(sine factors) which smoothens the long-distance behavior of the potential.

Besides the aforementioned situation which indicates the possible
occurrence of dangerous infrared singularities we would like to stress
that, in fact,
radiative corrections bring new infrared divergences which at higher
order lead to the breakdown of the $1/N$ expansion.
The crucial point
of the calculation is provided by the corrections to the $\varphi $
field two point function whose contributions are again
given by the graphs on the Figs. \ref{Fig2} and \ref{Fig2a} 
(we may have other diagrams containing the mixed propagator but these are
nonplanar diagrams without IR or UV divergences). 
Let us first examine those contributions for the situation {\underline 1} listed above.
  Unless for the graph
\ref{Fig2a}b which is still planar, now there are trigonometric
factors that deserve special consideration.   

In the Landau gauge ($C\rightarrow \infty$), in which we have chosen to work, the
transversality property of the gauge propagator produces a reduction
of the degree of divergence of graph \ref{Fig2a}a by two units. 
Indeed the amplitude for
the graph \ref{Fig2a}a turns out to be

\vspace*{0.5cm}

\begin{eqnarray}
\int \frac{d^{3}k}{(2\pi )^{3}}\frac{(2p+k)_{\mu }\, (2p+k)_{\nu }}{(p+k)^{2}
-m^{2}}&&\!\!\Delta ^{\mu \nu }(k)\sin ^{2}(k\wedge p)\nonumber \\
&&
=4p_{\mu }p_{\nu }\int 
\frac{d^{3}k}{(2\pi )^{3}}\frac{1}{(p+k)^{2}-m^{2}}\Delta ^{\mu \nu }(k)
\sin ^{2}(k\wedge p).
\end{eqnarray}

\noindent
The  ultraviolet (logarithmic) divergence of this expression must be removed
by an adequate counterterm; no infrared divergence appears because
the sine factors improve the behavior of the integrand for small momenta.
However, graph \ref{Fig2a}c has a leading contribution which, for
high loop momenta, behaves as

\begin{equation}
\int \frac{d^{3}k}{(2\pi )^{3}}\cos (2k\wedge p)\frac{1}{\sqrt{{k}^{2}}}\label{11}
\end{equation}

\noindent 
and is quadratically divergent as $p$ goes to zero. The multiple insertions of this graph into a larger graph
leads to nonintegrable singularities which destroy the $1/N$
expansion. At this point we may wonder if this result could not be
modified by another choice for the trilinear interaction among the
$\lambda$ and $\varphi$ fields. In fact, the choice (\ref{30}) as
the interaction part involving the $\lambda$ field
introduces a sine factor at the trilinear vertex as it already happens
with the   $A_\mu A^\nu \varphi \varphi^\dag $ vertex.  If this is
done then diagram of Fig. \ref{Fig2a}b would have also a nonplanar
part which asymptotically is similar to (\ref{11}). However, the numerical factors do
not match and no cancellation could take place.

\section{Concluding remarks}

In this work we focused on the construction of a consistent extension
of the $CP^{N-1}$ model to the noncommutative space. As we have seen, there
are various possible extensions which depend on the way the fields
transform under the gauge group. In all situations, we have chosen
the auxiliary field $\lambda $ as belonging to the adjoint representation.
For the $\varphi$ field belonging to the fundamental representation 
  this prescription automatically prevents
the appearance of a mixed $<T\lambda A_\mu>$ propagator. 
In  fact the possibility envisaged in Eq. (\ref{5})
leads to a nonvanishing two-point proper function of the $A_{\mu }$
and $\lambda $ fields given by

\begin{equation}
\Gamma _{A_{\mu }\, \lambda }(p)=-N\int \frac{d^{3}k}{(2\pi )^{3}}\, 
\frac{(2k+p)_{\mu }}{(k^{2}-m^{2})[(k+p)^{2}-m^{2}]}{\textrm{e}}^{-i2k
\wedge p}=\frac{iN\tilde{p_{\mu }}}{8\pi 
\sqrt{-{\tilde{p}}^{2}}}\int_{0}^{1}dx{\rm e}^{-M\sqrt{-{\tilde {p}^2}}}.
\end{equation}

\noindent
For the $\varphi$ field belonging to the
adjoint representation, the mixing of the $\lambda$ and $A_\mu$ fields will occur unless if the constraint Lagrangian is
as in Eq. (\ref{30}).

 Up
to the leading nontrivial order of $1/N$, all dangerous IR divergences
were shown to cancel if the basic $\varphi $ field belongs to the
left representation of the gauge group. We also proved that
the ultraviolet divergences may be absorbed into counterterms which
preserve the form of the original Lagrangian. Therefore gauge invariance
is maintained but this occurs in a way different from the commutative case.
Indeed, in the commutative setting all the counterterms coefficients
$b$, $B$ and $C$ defined in (\ref{count}) are equal and the $A_\mu$
field is not renormalized. In the present situation, however, the
 $A_\mu$ field gets renormalized and, although innocuous, a quadrilinear
vertex $ A_\mu A^\nu\varphi\varphi^{\dag}$  counterterm occurs.

An entirely different picture is found if the basic field belongs to
the adjoint representation. First, the graphical identities
characteristics of the commutative model are no longer
valid. Nonplanarity occurs already at the leading order of $1/N$ and
the number of diagrams to be analyzed increases significantly. Also,
due to the presence of (Moyal) commutators, the $A_\mu$ field formally
decouples from the theory in the commutative limit. However, in this limit the Green functions are singular
and the limit does not seem to exist (this is, of course, also true
if the $\varphi$ field is in the fundamental
representation). Dangerous infrared divergences occur both in the
gauge sector and in the radiative corrections to the two point
function of the $\varphi$ field. 

Because of the noncommutativity of the Moyal product
there is a two parameter family of interaction terms
containing the auxiliary $\lambda$ field. However, for no choice it is
possible to cancel the divergences.
Actually, the existence  of IR/UV mixing suggests that to achieve consistency
further extensions of the model should be investigated. This issue
is the object of our next study where the inclusion of fermionic matter
fields will be investigated.  

\section{Acknowledgments}

This work was partially supported by Funda\c{c}\~{a}o de Amparo
\`{a} Pesquisa do Estado de S\~{a}o Paulo (FAPESP) and Conselho
Nacional de Desenvolvimento Cient\'{\i}fico e Tecnol\'{o}gico (CNPq).

\appendix
\section{}

\begin{figure}
\begin{center}
\includegraphics{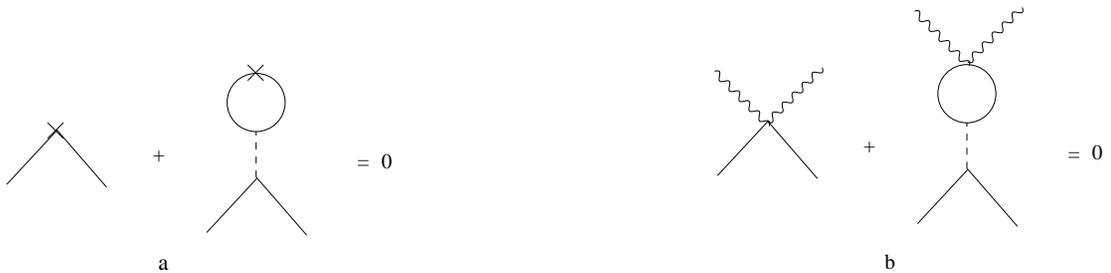}
\end{center}
\caption{Graphical identities for the $CP^{N-1}$ model.}
\label{Fig1a}
\end{figure}

\begin{figure}
\begin{center}\includegraphics{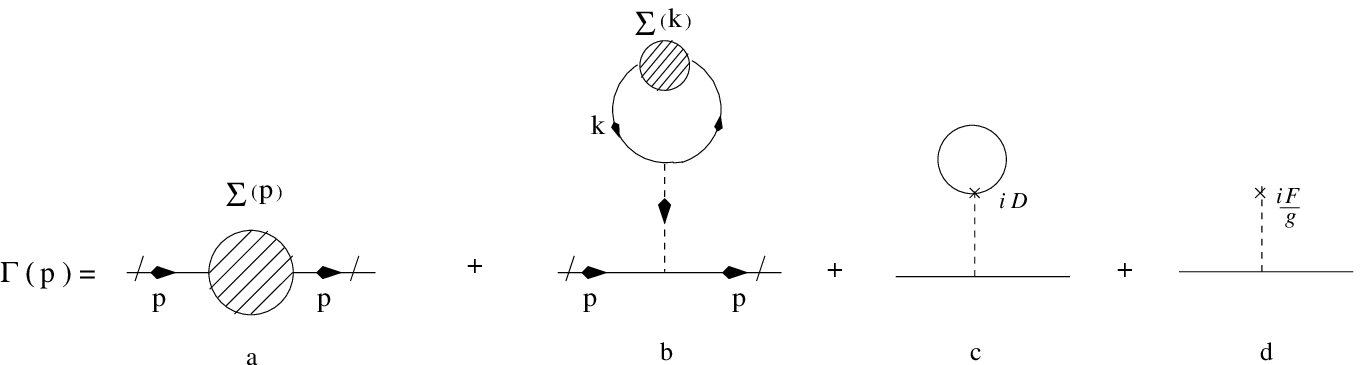}\end{center}
\caption{Graphical structure of the $\varphi$ field two point function.The
hachured bubble represents diagrams that are 1PI with respect to all fields.}
\label{Fig2}
\end{figure}

\begin{figure}
\begin{center}\includegraphics{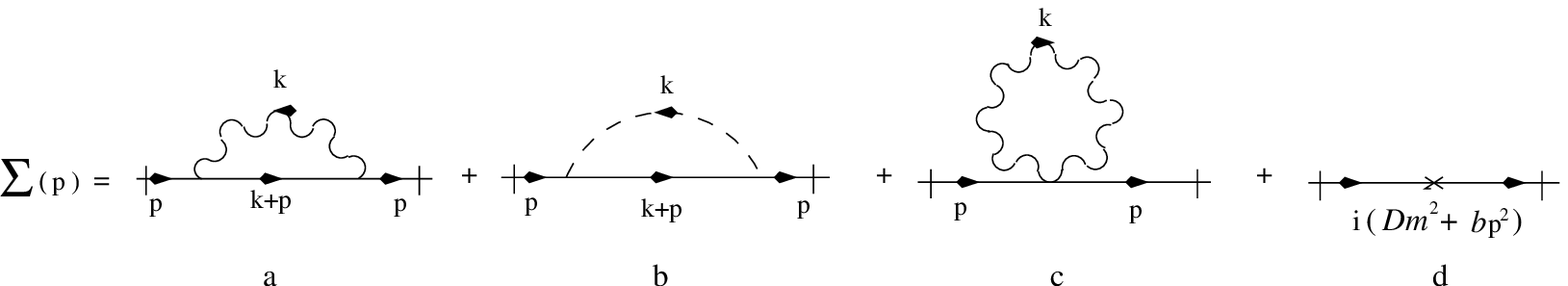}\end{center}
\caption{Subleading contributions to the $\varphi $ propagator.}
\label{Fig2a}
\end{figure}

\begin{figure}
\begin{center}\includegraphics{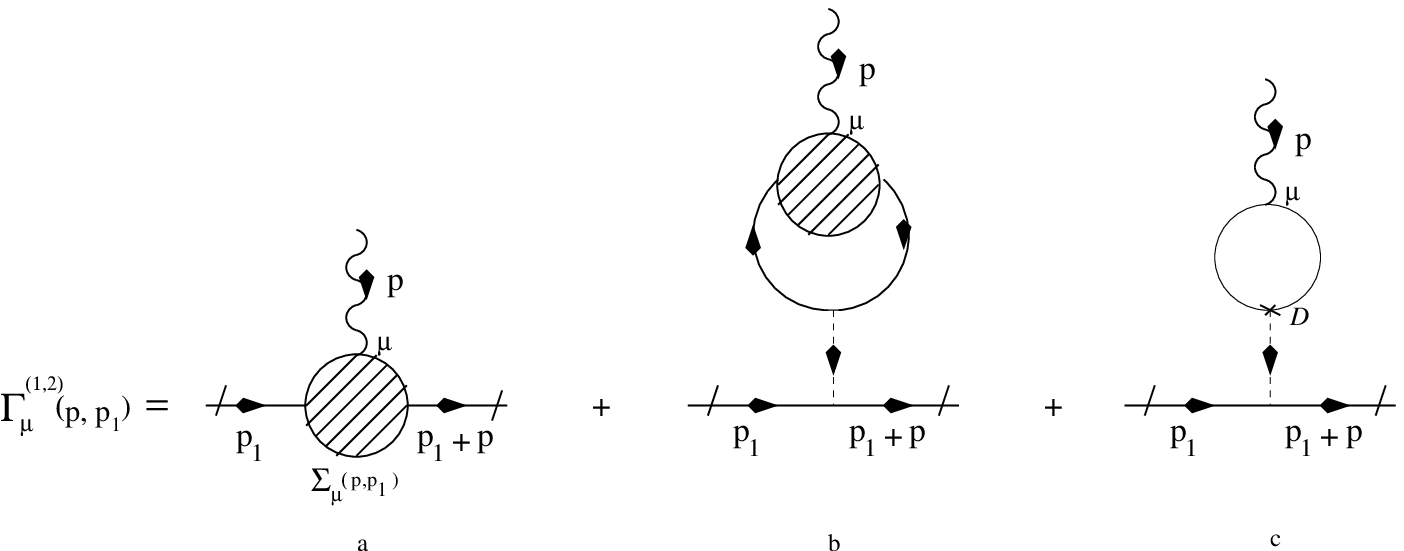}\end{center}
\caption{General structure of the three-point vertex function of 
the $A_\mu $ and $\varphi $ fields.}
\label{Fig3a}
\end{figure}

\begin{figure}
\begin{center}\includegraphics{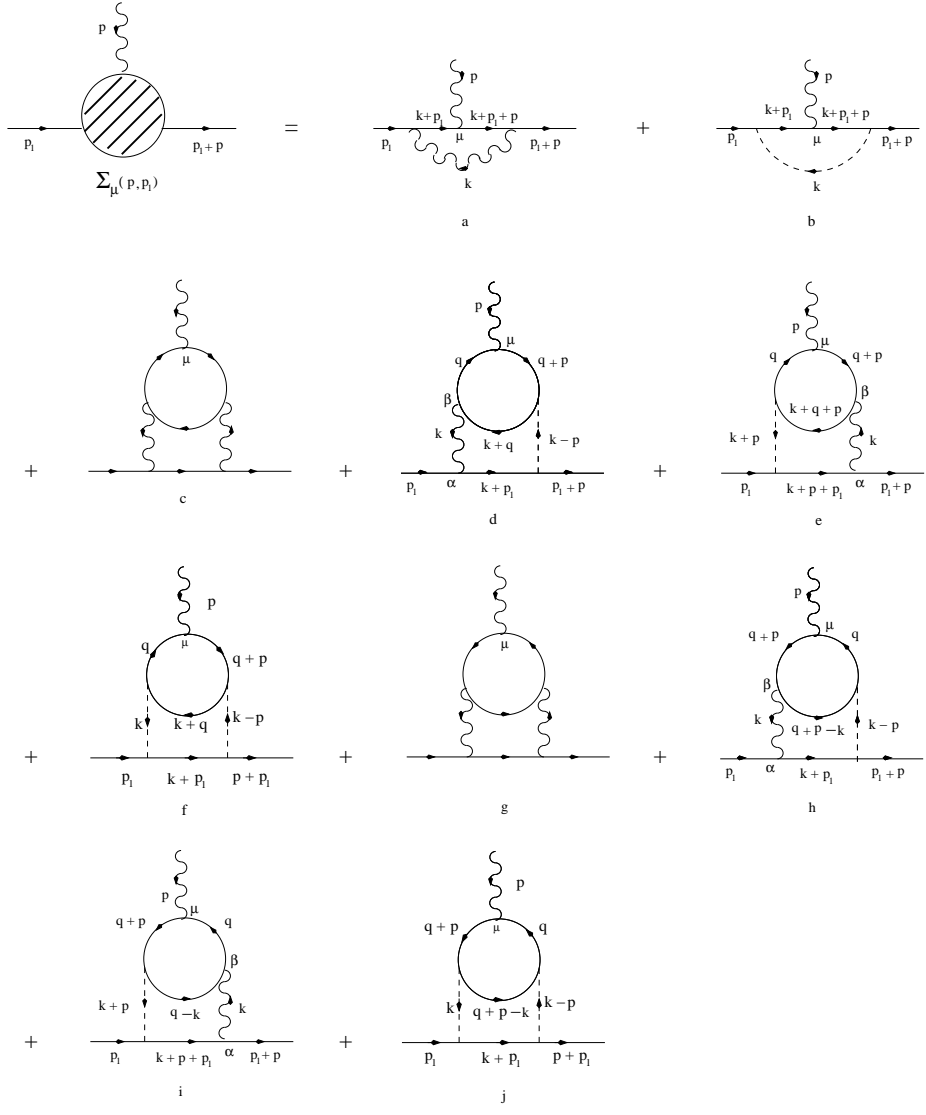}\end{center}
\caption{Three-point function of the $A_\mu $ and $\varphi $ fields.}
\label{Fig3}
\end{figure}

\begin{figure}
\begin{center}\includegraphics{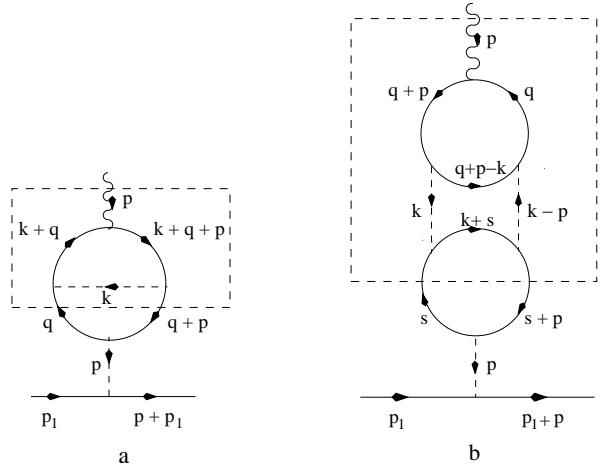}\end{center}
\caption{Compensating diagrams for the graphs \ref{Fig3}b and \ref{Fig3}j}
\label{Fig3b}
\end{figure}

\begin{figure}
\begin{center}\includegraphics{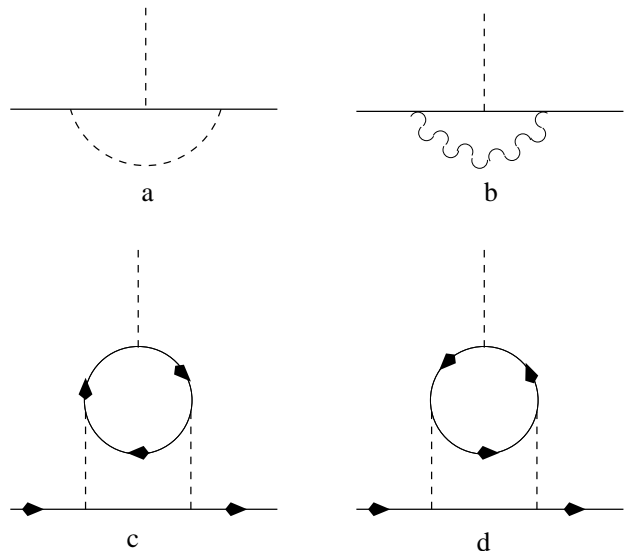}\end{center}
\caption{Three-point function of the $\lambda $ and $\varphi $ fields.}
\label{Fig31}
\end{figure}

\begin{figure}
\begin{center}\includegraphics{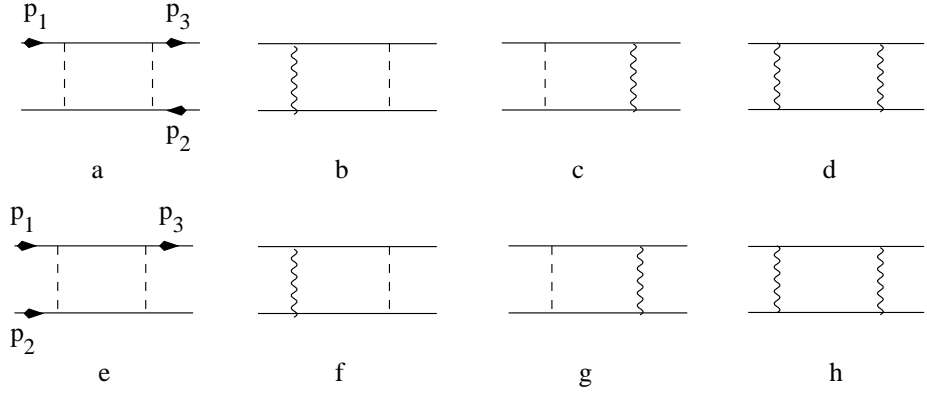}\end{center}
\caption{Four-point function of the $\varphi $ field.}
\label{Fig4}
\end{figure}

\begin{figure}
\begin{center}
\includegraphics{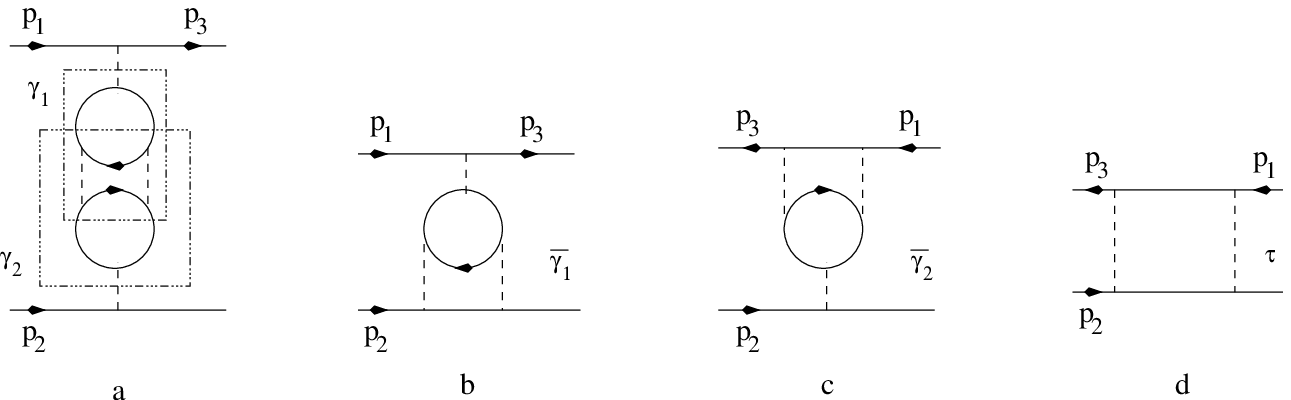}
\end{center}
\caption{Illustration of the mechanism for the complete cancellation of the
UV divergences in the four point function of
the $\varphi $ field. The first graph ($G$) contains $\gamma_1$, $\gamma_2$
 and $\tau$ as subgraphs.}
\label{Fig5}
\end{figure}

In this Appendix we shall demonstrate some results concerning the
ultraviolet behavior of the noncommutative $CP^{N-1}$ model when the
basic $\varphi$ field belongs to the left fundamental representation
of the gauge group. They are:

a. The subleading contributions to the $\lambda$ and $A_\mu$
propagators, which are explicit in (\ref{9}) and (\ref{51}) are
irrelevant as far as the ultraviolet divergences are concerned. In
fact, the only case that requires special attention is the two point
function of the $\varphi$ field which becomes at most linearly
divergent if the subleading contribution for the auxiliary fields is
used.  Actually, up the order that we have considered it only occurs 
in the diagram of Fig. \ref{Fig2a}b.  Replacing this contribution in
Figs. \ref{Fig2} and \ref{Fig2a} one sees that the would be linear
divergences cancel among themselves. The next subdivergence which is only
logarithmic vanishes due to Lorentz covariance.

b. Contributions containing the mass counterterm cancel pairwise.
This result follows straightforwardly from the graphical identity
depicted in Fig. \ref{Fig1a}a where the special vertex stands for
the mass counterterm insertion.

c. Finally we will exemplify how the complete cancellation of
ultraviolet divergences takes place in the case where several (sub)
diagrams are involved, as in the three-loops diagram $G$ of
Fig. \ref{Fig5}. The ultraviolet divergent subgraphs of $G$ are: $
\gamma_1, \gamma_2$ and $\tau$; they also occur as (sub)graphs of the
diagrams $ \bar\gamma_1, \bar\gamma_2$ and $\tau$ as shown in
Fig. \ref{Fig5} (in spite of having different number of loops they are
of the same order in $1/N$). These diagrams are all planar and
have the same noncommutative phase  ($\exp i[p_1\wedge p_2-p_1\wedge p_3+p_2\wedge p_3]$)  which therefore factorizes in their sum.
Notice that $\gamma_1$ and $\gamma_2$
are overlapping and that both contain $\tau$ as a subgraph. In the
BPHZ scheme the relevant $G$-forests are $\not\! 0, \gamma_1,
\gamma_2, \tau, \{\gamma_1,\tau\}$ and $\{\gamma_2 ,\tau\}$. The BPHZ
subtracted amplitude associated to the graph $G$ is therefore

\begin{equation}
R_G= I_G- I_{G/\gamma_1}t^{0}_{\gamma_1} I_{\gamma_1}- I_{G/\gamma_2}
t^{0}_{\gamma_2} I_{\gamma_2}- I_{G/\tau}t^{1}_{\tau} I_{\tau}+
I_{G/\gamma_1}t^{0}_{\gamma_1}I_{\gamma_1/\tau}t^{1}_{\tau} I_{\tau}+
I_{G/\gamma_2}t^{0}_{\gamma_2}I_{\gamma_2/\tau}t^{1}_{\tau} I_{\tau},
\end{equation}

\noindent
where $I_G$ denotes the unsubtracted amplitude associated with the
graph G; as it is usual, $I_{G/\gamma}$ is the amplitude associated
to the reduced graph $G/\gamma$ obtained by contracting the
subgraph $\gamma$ of $G$ to a point. For a generic diagram $\gamma$, 
$t_\gamma$ is defined as the Taylor operator on the external independent momenta of $\gamma$ with the proviso that it does not act on the noncommutative
phase factor.    
 Similarly, the BPHZ subtracted amplitudes for the graphs
$\bar\gamma_1$, $\bar\gamma_2$ and $\tau$ are

\begin{eqnarray}
R_{\bar\gamma_1}&=& I_{\bar\gamma_1}- 
I_{\bar \gamma_1/\gamma_1}t^{0}_{\gamma_1} I_{\gamma_1} - I_{\bar\gamma_1/\tau}
t^{1}_{\tau} I_\tau+ I_{\bar \gamma_1/\gamma_1}t^{0}_{\gamma_1}
I_{\gamma_1/\tau}t^{1}_{\tau} I_{\tau},\\
R_{\bar\gamma_2}&=& I_{\bar \gamma_2}-I_{\bar \gamma_2/\gamma_2}
t^{0}_{\gamma_2} I_{\gamma_2} - I_{\bar\gamma_2/\tau}
t^{1}_{\tau} I_\tau+ I_{\bar \gamma_2/\gamma_2} t^{0}_{\gamma_2}
I_{\gamma_2/\tau}t^{1}_{\tau} I_{\tau}
\end{eqnarray}

\noindent
and

\begin{equation}
R_\tau= I_\tau- t^{1}_{\tau} I_\tau
\end{equation}

Noticing now that, as consequence of the graphical identity in
Fig. \ref{Fig1a}a, $I_{G/\gamma_1}=I_{G/\gamma_2}=-I_{\bar
\gamma_1/\gamma_1}=-I_{\bar \gamma_2/\gamma_2}$ and that
$I_{G/\tau}=-I_{\bar\gamma_1/\tau}=-I_{\bar\gamma_2/\tau}= $ trivial
four vertex, we see that when adding the above contributions all the
subtraction terms cancel. We stress that the cancellation occurs for
all subtractions including those associated to the last subtraction
for the linearly divergent diagram $\tau$ (in this case, to the reduced vertex
associated to the contraction of $\tau$ to a point it is assigned a linear
polynomial in the external momenta of $\tau$).
This proves that the sum of the unsubtracted
diagrams is finite.

If the charge flow in the  upper and lower loops of Fig. \ref{Fig5} are in opposite directions the corresponding diagrams are nonplanar. They still have the
same phase factor but it depends on the loop momentum of the $\tau$ diagram.
Individually they present a linear infrared divergence which nonetheless
is cancelled whenever they are added. This is most easily seen by 
factorizing the noncommutative phases and then
Taylor
expanding the remaining of the $\tau$'s integrand up to first order in the independent  external momenta of that graph.

\newpage
 

%\begin{figure}
%\begin{center}
%\includegraphics{figura11.eps}\end{center}
%\caption{Some additional graphs contributing to the four point function of
%the $\varphi $ field.}
%\label{Fig5}
%\end{figure}

\vspace*{-4.0cm}

\begin{thebibliography}{10}
\bibitem{footnote1} Logarithmic infrared divergences resulting from
the IR/UV mixing may occur but these are not dangerous because they
are integrable. As discussed in \cite{Minwalla}, unlike the case of
logarithmic divergences, multiple insertions of quadratic or linear
infrared divergent graphs into a larger graph will produce
nonintegrable singularities.
\end{thebibliography}
\end{document}